\documentclass[journal]{IEEEtran}
\usepackage{amsmath,amsfonts,amsthm}
\usepackage{stfloats}
\usepackage{enumitem}
\usepackage{verbatim}
\usepackage[misc]{ifsym}
\usepackage{graphicx}
\usepackage{amsmath}
\usepackage{mathtools}
\usepackage{algorithmic}
\usepackage{algorithm}
\usepackage[table,xcdraw]{xcolor}
\usepackage{orcidlink}
\hypersetup{
    hidelinks
}

\DeclareMathOperator*{\argmax}{argmax}
\DeclareMathOperator*{\argmin}{argmin}

\begin{document}

\title{Underdetermined Library-aided Impedance\\ Estimation with Terminal Smart Meter Data}

\author{ \IEEEauthorblockN{Federico Rosato \orcidlink{0000-0001-7101-8273}},
        \IEEEauthorblockN{Lorenzo Nespoli \orcidlink{0000-0001-9710-8517},
        \IEEEauthorblockN{Vasco Medici
        \orcidlink{0000-0001-5684-4972}}
        }
\thanks{This work has been submitted to an IEEE journal for possible publication. Copyright may be transferred without notice, after which this version may no longer be accessible.}
\thanks{The authors are with the University of Applied Sciences and Arts of Southern Switzerland (SUPSI).}
\thanks{ \Letter \ \{federico.rosato, lorenzo.nespoli, vasco.medici\}@supsi.ch}
}

\markboth{\ }
{Federico Rosato: Underdetermined Library-aided Impedance Estimation with Terminal Smart Meter Data} 

\maketitle


\begin{abstract}
Smart meters provide relevant information for impedance identification, but they lack global phase alignment and internal network nodes are often unobserved. A few methods for this setting were developed, but they have requirements on data correlation and/or network topology. In this paper, we offer a unifying view of data- and structure-driven identifiability issues, and use this groundwork to propose a method for underdetermined impedance identification. The method can handle intrinsically ambiguous topologies and data; its output is not forcedly a single estimate, but instead a collection of data-compatible impedance assignments. It uses a library of plausible commercial cable types as a prior to refine the solutions, and we show how it can support topology identification workflows built around known georeferenced joints without degree guarantees. The method depends on a small number of non-sensitive parameters and achieves high identification performance on a sizeable benchmark case even with low-size injection/voltage datasets. We identify key steps that can be accelerated via GPU-based parallelization. Finally, we assess the tolerance of the identification to noisy input.

\end{abstract}


\section{Introduction}

In low voltage distribution networks, accurate and complete impedance information is rarely available, but it is increasingly important for evaluation of grid impact of distributed generation and electrical vehicles, distributed control, et cetera. Smart meters offer valuable data at the delivery points (active and reactive power, voltage magnitude), but only in local phase reference, and no measurements are usually made at internal network nodes. This information is in general insufficient for unique and exact parameter estimation, but the setting remains attractive because it does not require the installation of additional instrumentation beyond the smart meter themselves. On the other hand, in real scenarios it is often possible to obtain a prior distribution on the electrical properties per unit length of the wires used on field (which we call a \textit{library}). This piece of information, which has seen limited use \cite{Guo2022-ax}, can be leveraged to drive solutions towards standardized or otherwise known cable types. 

In the literature there are a number of works relevant to the smart meter data setting. The 2024 general review on applications of smart meter data by Athanasiadis \cite{Athanasiadis2024-go} provides a good overview of topology/impedance estimation. The authors of \cite{Zhang2021-os} estimate the admittance matrix without topology but with measurements at all nodes. In \cite{Han2015-on} and \cite{Peppanen2015-bv}, impedance estimation is performed as an intermediate step in specific applications, and is limited to the considered linear configurations. \cite{Wang2020-dr} studies a condition where all buses are either metered or separated from a metered bus by a service cable whose impedance is known in advance. Iakovlev et al. \cite{Iakovlev2018-cz} present a fast method for impedance estimation that optionally uses an r/x prior for enhancing the well-posedness of the regression, but it needs measurement at all nodes. Another important work is \cite{Park2020-if}, in which topology and impedance estimation are performed with only terminal node measurements; however, the identification in the presence of latent nodes hinges on statistical injection independence among terminal nodes. Furthermore, cases with latent nodes of degree $2$ are not considered, as their solution is intrinsically non-unique \cite{Deka2016-jd}. The works arguably closest to the present paper are \cite{Tong2022-el} and the recent \cite{Guo2025-hp} by Guo et al. Both of these papers tackle the same setting, and instead of statistical cross-influence of nodal injection and voltages, they use leaf-to-root backward induction with voltage and injection snapshots to treat latent nodes, as is done in the present paper. However, potential ill-conditioning issues for the underlying regression/optimization problems are not explicitly considered, and in order to offer directly solvable formulations, both of these papers exclude degree 2 internal nodes to avoid ambiguity. Retaining degree-2 nodes is useful whenever the objective is asset-level feeder reconstruction rather than mere electrical equivalencing. Although collapsing consecutive segments is often sufficient if one only seeks to reproduce terminal P, Q, and V behaviour, it becomes restrictive in applications such as digital-twin construction, completion of missing asset registries, branch-specific reinforcement planning, maintenance prioritization, and segment-level ampacity assessment. Cable-type attribution along GIS-mapped segments is a further application we put a strong focus on. In the last part of the paper, we envision it as a building block for estimating the likelihood of topology proposals over GIS-located joints. In these settings, the total impedance of an ambiguous chain is not the only quantity of interest; it is also important to infer which physically plausible allocation of impedance across the constituent segments is consistent with the measurements and with the available conductor library. Preserving degree-2 nodes therefore makes it possible to represent structural ambiguity explicitly at the segment level, instead of absorbing it into a single equivalent branch. 
Summarizing, compared to the existing literature, this paper:
\begin{enumerate}[leftmargin=*]
    \item Introduces an identifiability analysis based on LinDistFlow that deals with intrinsically ambiguous networks and clarifies many identifiability issues from a unified perspective;
    \item With this theoretical underpinning, proposes an underdetermined identification method that can output a range of data-compatible solutions instead of a one-shot estimate;
    \item Makes use of a (often easy to get) prior on per-meter impedances to mitigate identifiability problems;
    \item Does not fail silently under problematic data.
\end{enumerate}

\subsection{Structure of the paper}

\begin{table*}[b]
\centering
\caption{Summary of parameters.}
\begin{tabular}{cll}
\rowcolor[HTML]{EFEFEF} 
{\color[HTML]{000000} \textbf{Symbol}} & {\color[HTML]{000000} \textbf{Description}} & {\color[HTML]{000000} \textbf{Trade-off when increasing}} \\
$m$       & Number of samples                 & Runtime increases. Better coverage of feasible space.                 \\
$\kappa$    & Maximum residual coefficient ($>1$)		         & Better capture of solution range, but increases their admissible infeasibility.  \\
$\lambda$ & PGD learning rate                                & PGD may oscillate or diverge, but converges faster if it does at all. \\
$\rho$ & PGD penalty coefficient                                & Final solutions closer to the prior, but infeasibility might increase. \\
$m'$      & Final number of proposed solutions                & User preference.                                                      \\
$K$       & Number of neighbors for Facility Location kernel & Better quality of thinning, but runtime and memory usage increase.  \\     
$\{m,b,r,x\}_{\{\mathrm{hi},\mathrm{lo}\}}$ & Impedance polytope absolute bounds & Better capture of library deviations, but looser range.

\end{tabular}
\label{tab:params}
\end{table*}

\begin{figure}[hbt!]
    \centering
    \includegraphics[trim={1.2cm 1.5cm 1.2cm 1cm},clip,width=\columnwidth]{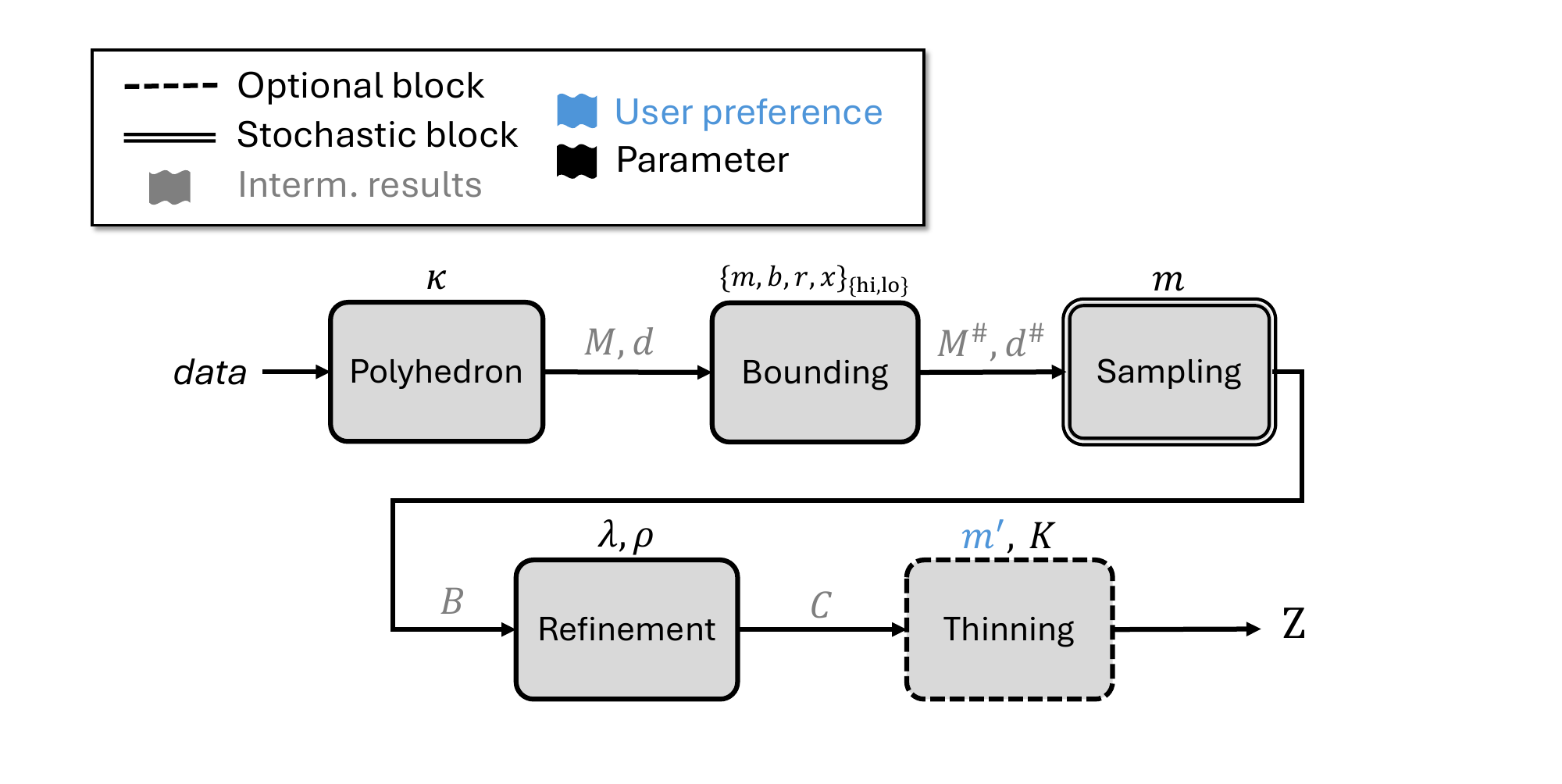}
    \caption{Block scheme overview of the method.}
    \label{fig:block_scheme}
\end{figure}

The remainder of the paper is organized as follows. In section \ref{sec:setup}, we state precisely the problem setup we are tackling. Sections \ref{sec:struct} and \ref{sec:geometry} are the backbone of the paper, where we derive and analyze the feasible impedance polyhedron $\mathcal{Z}$ under the LinDistFlow regime. We deem this analysis is of broad interest, as it encompasses many identifiability issues from a unified geometric and algebraic perspective. Furthermore, it provides the theoretical underpinning for an associated identification pipeline, whose steps are described starting from Section \ref{sec:gampl}, where we present library-based bounding of $\mathcal{Z}$ to obtain polytope $\mathcal{Z}^\#$ and a toolchain to obtain samples from the latter. In Section \ref{sec:refinement}, the collection of samples is condensed around solutions in which the branch impedances are as close as possible to the library values while encouraging continued compatibility with the data through a penalty term, in a process we call refinement. This is done via gradient descent with a tunable penalty term, to prevent excessive drifting into technically infeasible regions in case the library prior or length estimates are not considered strongly reliable. In Section \ref{sec:thinning}, an optional step based on submodular optimization is proposed as a way to select a maximally representative subset of solutions, much smaller than the full sampling. The identification pipeline is schematized in Figure \ref{fig:block_scheme}, with a summary of the necessary parameters in Table \ref{tab:params}.

In Section \ref{sec:i3e} we test the identification approach on the IEEE European Low Voltage Test Feeder, evaluating the offered solution range for this dataset against the known ground truth. Finally, in Section \ref{sec:noise}, we analyze the tolerance of the method to noisy data.

\section{Problem setup}\label{sec:setup}
We consider a 1-phase or balanced 3-phase radial electric distribution feeder, represented by a tree graph $\mathcal{T} = (\mathcal{N}, \mathcal{E})$. Nodes are named by consecutive integers $n\in\{0,\dots\}$. Throughout the paper, edges are either named by an index $e$ in the same fashion, or addressed by the nodes they join in parentheses: $(n_1,n_2)$. The tree is rooted at node $0$, which corresponds to the distribution cabinet for the feeder. The subset of leaf nodes is $\mathcal{L} \subset \mathcal{N}$. The cardinality of the edge set for a tree is $|\mathcal{E}|=|\mathcal{N}|-1$.

We assume that the length $\ell_e$ of the conductor of branch $e\in \mathcal{E}$ is known, at least approximately.

The identification method is based on smart meter data. We assume to have, for a set of datapoints $\{1, \dots, t, \dots, T\}$, the nodal data summarized in Table \ref{tab:data}:

\begin{table}[H]
\caption{Required smart meter data for \break the impedance identification method}
\centering
\begin{tabular}{|l|c|c|}
\hline
\rowcolor[HTML]{EFEFEF} 
\textbf{Data}     & \textbf{Symbol} & \textbf{Nodes} \\ \hline
Active power      & $P_{t,n}$       & $n \in \mathcal{N}$              \\ \hline
Reactive Power    & $Q_{t,n}$       & $n \in \mathcal{N}$              \\ \hline
Voltage magnitude & $v_{t,n}$       & $n \in \mathcal{L} \cup \{0\}$ \\ \hline
\end{tabular}
\label{tab:data}
\end{table}

Voltage magnitude is needed only at the leaves and the root. Powers $P_{t,n}$ and $Q_{t,n}$ are required at all nodes, but the natural setting for the method is the one where the inner nodes are passive (unloaded) and correspond to joints, bifurcations. In this case, the actual power measurements are made for $n \in \mathcal{L}$, and  $P_{t,n}=Q_{t,n}=0$ for $n \notin \mathcal{L}$.
In fact, if inner node $n \in \mathcal{N} \setminus \mathcal{L}$ is loaded and metered, a voltage measurement shall be available as well. In that case, the tree network can be split at $n$, obtaining subtree $\mathcal{T}_n$ rooted at $n$. $\mathcal{T}_n$ can be solved directly, and the aggregated power at $n$ can be used as a load power to solve $\mathcal{T} \setminus \mathcal{T}_n$. The reasoning can be applied recursively until only trees with unloaded (or no) inner nodes remain.

We assume that the data associated with each $t$ are temporally synchronized. No global phase measurement is required.


\subsection{Wire library}

In this paper, we will use the term \textit{library} to define a list of couples of resistance, reactance per unit length, each associated with the nominal characteristics of a candidate type of cable for network branch $e$.

\begin{equation*}
    \begin{aligned}
         \mathbf{c}_e = \{c_{e,1}, \dots,c_{e,k}\} = \{(r_{e,1} +i x_{e,1}), \dots, (r_{e,k}+i x_{e,k})\}&  \\
         \text{for some } k \in \mathbb{N}&
    \end{aligned}
\end{equation*}

Such libraries can be obtained by DSOs in real studies by leveraging a history of procurement, knowledge of the standard sizes used on field, etc. The library is specified per branch, and to facilitate the identification task it should include the least possible number of candidates for that branch among which the actual type used is present. In the limit, the libraries each include only one type, in which case the identification task is trivial.

In any identification procedure using a library, it would be difficult to distinguish between two species of cables with indistinguishable impedances per length; this is out of the scope of the paper, and we will assume that only well-distinct points are included in each branch library.

\begin{equation}
    \begin{aligned}
        \frac{|c_{e,i}-c_{e,j}|}{\max(|c_{e,i}|,|c_{e,j}|)} \ge \varepsilon \;\; \forall i, j \in \{1,\dots,k\}&
\\ \text{for some fixed}\; \varepsilon>0&
    \end{aligned}
\end{equation}

\begin{figure}[hbt!]
    \centering
    \includegraphics[trim={0cm 0.7cm 0cm 0.7cm},width=\columnwidth]{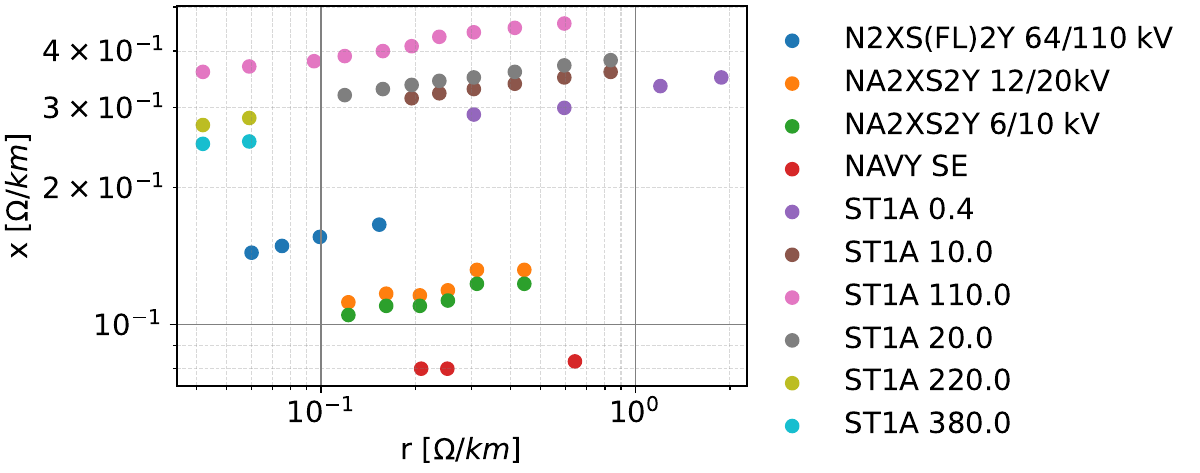}
    \caption{Libraries derived from the Pandapower standard line types. The graph is in log scale on both axes.}
    \label{fig:pandapower_catalogs}
\end{figure}

This poses no problem in the basic identification case, in which the library is derived from a collection of cables of the same type but different sections. We wish to stress once more that individual libraries can be specified per branch (as an outstanding example, one may wish to use distinct impedance libraries for underground and aerial branches); similar impedances across libraries are not a problem. This is exemplified in Figure \ref{fig:pandapower_catalogs}, where various libraries of cables included in the standard c of the simulation package Pandapower \cite{Thurner2018-ts} are shown. 

\section{Derivation of feasible impedance polyhedron} \label{sec:struct}
By adopting the LinDistFlow \cite{Baran1989-fb} approximation, it is possible to perform a fit of the power model to the available data. LinDistFlow is a linearization of the power flow equations around the flat voltage profile. It introduces aggregated branch powers for edge $(p,n) \in \mathcal{E}$:

\begin{align}
    \mathbf{P}_{t,(p,n)} &\approx P_{t,n} + \sum_{s \in \{s|(n,s)\in\mathcal{E} \}} \mathbf{P}_{t,(n,s)}\\
    \mathbf{Q}_{t,(p,n)} &\approx Q_{t,n} + \sum_{s \in \{s|(n,s)\in\mathcal{E} \}} \mathbf{Q}_{t,(n,s)}
    \label{eq:lindistflow_powers}
\end{align}

\noindent then, the difference in square voltage magnitudes is approximated by: 
\begin{equation}
    \mathbf{v}_{t,p}^2 - \mathbf{v}_{t,n}^2 \approx 2\big(r_{p,n}\mathbf{P}_{t,(p,n)} + x_{p,n}\mathbf{Q}_{t,(p,n)}\big)
    \label{eq:v2drop}
\end{equation}

If we define an appropriate incidence matrix $A$ in the following way:
\begin{equation}
A_{n,e} = 
    \begin{cases}
    1 \leftarrow e \in \text{path}_{\mathcal{T}}(0,n) \\
    0 \leftarrow otherwise
    \end{cases}
\end{equation}

\noindent the predicted square voltages can be telescoped to the root slack bus with unitary voltage and compactly written in the following way:

\begin{align}
        \mathbf{v}^2 &= \mathbf{1}- 2\big( \mathbf{P} \odot (\mathbf{1}_T  \mathbf{r}) +\mathbf{Q} \odot (\mathbf{1}_T  \mathbf{x}) \big)A^{\top}
        \label{eq:compactpredvolt}
\end{align}

We call the flattened impedance vector $\mathbf{z}=[\mathbf{r}, \mathbf{x}]$. If we treat the injections as fixed parameters, the predicted square voltage just depends on the impedance vector: $\mathbf{v}^2=\mathbf{v}^2(\mathbf{z})$. We then cast an optimization problem representing the best-fit approximation of $\mathbf{z}$ to the datapoint $t$ in the following way:

\begin{equation}
\begin{aligned}
\min_{\mathbf{z},\Delta } \quad & \Delta\\
\textrm{s.t.} \quad & v^2_{t,n} - \Delta \leq  \mathbf{v}^2_{t,n}(\mathbf{z}) \leq v^2_{t,n} + \Delta \quad n\in \mathcal{L} 
\end{aligned} \label{eq:deltaopt}
\end{equation}

Solving this linear program yields an impedance estimate $\mathbf{z}^*$, and a minimum modeling error $\Delta^*$, representing the maximum portion of any measured voltage point that cannot be explained by LinDistFlow given injections $P$ and $Q$. Equivalently, if $\Delta^*=0$, the data is an exact LinDistFlow solution for some $\mathbf{z}$.

We are interested in the region $\mathcal{Z}$ that $\mathbf{z}$ occupies while remaining compatible with the data in the LinDistFlow sense up to a coefficient $\kappa$:

\begin{equation}
    \mathcal{Z} = \{\mathbf{z}\ |\ v^2_{t,n}- \kappa\Delta^* \leq \mathbf{v}^2_{t,n}(\mathbf{z}) \leq v^2_{t,n}+ \kappa\Delta^* \}
    \label{eq:implicit_polytope}
\end{equation}

This set is a polyhedron that can be expressed in half-space form by opportunely rearranging the inequalities of (\ref{eq:implicit_polytope}) and writing them explicitly according to (\ref{eq:compactpredvolt}). For each leaf $n\in \mathcal{L}$ we define matrix $M^{n} \in \mathbb{R}^{T\times 2|\mathcal{E}|}$ such that:

\begin{equation}
    M^{n}_{t,j} = 
    \begin{cases}
        2A_{n,e}\mathbf{P}_{t,e} &\leftarrow\ j=e \\
                2A_{n,e}\mathbf{Q}_{t,e} &\leftarrow\ j= |\mathcal{E}|+e
    \end{cases}
    \label{eq:m-matrix}
\end{equation}

Intuitively, each row $M^{n}_{t,\cdot}$ encodes the linear relationship between the relevant entries of $\mathbf{z}$ and the predicted voltage drop $\mathbf{v}^2_{t,n}$. We notice incidentally how, according to the particular incidence matrix $A$, $M^{n}$ can be -- and, in the typical case, is -- quite sparse. At this point, we can write two sets of inequalities for node $n$ (upper and lower bounds for all datapoints):

\begin{equation}
    \begin{bmatrix}
        -M^n \\
        M^n
    \end{bmatrix} \mathbf{z} \leq
    \begin{bmatrix}
        v^2_{\cdot,n} -\mathbf{1} + \kappa\Delta^*  \\
        \mathbf{1}-v^2_{\cdot,n} + \kappa\Delta^*
    \end{bmatrix}
    \label{eq:leaf_m}
\end{equation}

By stacking (\ref{eq:leaf_m}) for all leaves and labeling the stacked matrices and vectors $M \in \mathbb{R}^{2T|\mathcal{L}| \times 2|\mathcal{E}|}$ and $d \in \mathbb{R}^{2T|\mathcal{L}|}$ respectively, we get the half-space representation of the feasible impedance polyhedron $\mathcal{Z}$:

\begin{equation}
    \mathcal{Z} = \{\mathbf{z}\ |\ M\mathbf{z} \leq d \}
    \label{eq:halfspacerep}
\end{equation}

\section{The geometry of \texorpdfstring{$\mathcal{Z}$} as identifiability} \label{sec:geometry}

Due to $\mathbf{z}$ being sandwiched between lower and upper bounds that are tight up to $\kappa\Delta^*$ in (\ref{eq:leaf_m}), if $\Delta^* \approx0$, (\ref{eq:halfspacerep}) becomes an equality and $\mathcal{Z}$ is an affine space. If, further, $M$ is full rank, $\mathcal{Z}$ reduces to a single point, which corresponds to exact identification in the LinDistFlow regime; standard regression techniques are poised to produce well-conditioned and tight impedance estimates. If $\mathcal{Z}$ has a positive volume but $M$ remains well-conditioned, the polyhedron is bounded (a polytope). According to standard parameter estimation practice, the best individual estimate in terms of worst possible estimation error is the Chebyshev center:

\begin{equation}
\begin{aligned}
\mathbf{z}^0=\argmax_{\mathbf{z}}
\max_{r}\quad & r \\
\text{s.t.}\quad & M\mathbf{z} + r\,\|M\|_{2,\mathrm{row}} \le d\\
& r \ge 0
\end{aligned}
\end{equation}

Considerations on the deviation from ideal identifiability conditions are scattered throughout the literature; here we offer a review of identifiability issues through the unifying perspective of $\mathcal{Z}$.

\subsection{Rank deficiency due to data}

Bad conditioning or outright rank-deficiency of the matrix $M$ reflect inadequacies baked in the data. From a geometric standpoint, they correspond to singular directions in which $\mathcal{Z}$ extends indefinitely, making the corresponding linear composition of resistances and reactances non-identifiable. The assembly of the entries of $M$ in (\ref{eq:m-matrix}) reveals that failure to provide enough linearly independent injections collapses the row-rank of $M$. Equivalent rank issues in measurement matrices due to patterned injections are noted in \cite{Stanojev2023-gu}. Power ratios constantly very close to $1$ or $0$ are another well-known issue \cite{Grammatikos2025-mx}. At these extremes $\mathbf{Q}$, respectively $\mathbf{P}$, vanish. The right or left half of the $M$ matrix are consequently erased, halving the rank and making the corresponding half of $\mathbf{z}$ completely free from the constraints. The other half -- either $\mathbf{r}$ or $\mathbf{x}$ -- remains otherwise identifiable. A related phenomenon is experienced for any other constant power factor $\tan(\phi)$: in that case, inspection of (\ref{eq:m-matrix}) reveals that the left and right halves of $M$ are a multiple of one another. For brevity, let us analyze the case of negligible $\Delta^*$. If

\begin{equation}M=\bigl[M_{\text{half}}\;\;\tan(\phi) M_{\text{half}}\bigr]
\end{equation}

\noindent then (\ref{eq:halfspacerep}) becomes

\begin{equation}
    M_{\mathrm{half}}(\mathbf{r}+ \tan(\phi) \mathbf{x} ) \approx d
\end{equation}
\noindent in which $\mathbf{r}$ and $\mathbf{x}$ can be freely modulated in proportions governed by $\tan(\phi)$ without changing the solution. This echoes (\ref{eq:v2drop}), in which the same voltage magnitude drop can be obtained with the same aggregate powers by modulating $r_{p,n}$ and $x_{p,n}$ through the power factor. In practice, power factor induced identifiability issues can be settled with a library or a r/x prior, which anchors the proportions.

Other, subtler phenomena can impair identifiability in our setting. For example, the data might ``conspire" to always yield the same aggregated powers in upstream branches. However, this is not an inherent danger in the electrical distribution setting.

\subsection{Rank deficiency due to topology}

The issue with degree-2 nodes (\cite{Park2018-gq, Park2020-if, Deka2016-jd, Park2018-uj, Yuan2023-vg}) can be clearly appreciated as well. Two edges joined by a degree-2 node with no injection carry the same aggregated branch powers according to (\ref{eq:lindistflow_powers}), and the columns of $M$ corresponding to those edges will be identical, whatever the excitation and the amount of data. This, again, is made clear by (\ref{eq:m-matrix}). This immediately reduces the column-rank; on the other hand, if two such columns are substituted by a single one, and the corresponding entries of $\mathbf{z}$ are joined, we recover a full-rank identifiable system with a unified branch carrying the same aggregated power. This is akin to say that, albeit the individual branches are not identifiable, the \textit{sum} of their impedances is.

\subsection{Noise}

Noise is another broad and very relevant theme. In \cite{Grammatikos2025-mx}, the authors note that noise (which, notably for applications, can come in the form of time-averaging of measurements, as opposed to instantaneous snapshots) correspond to data not satisfying power flow equations. In our bound setting (\ref{eq:deltaopt}), the non-satisfaction is quantified in the LinDistFlow sense through $\Delta^*$, which in turn directly inflates $d$. In geometrical terms, the volume of the polytope is expanded by pushing each facet orthogonally by $\kappa\Delta^*$, increasing the set of compatible $\mathbf{z}$. The exact sensitivity $\partial(\mathrm{vol}(\mathcal{Z}))/\partial\Delta^*$ depends on $M$ and $d$ combinatorially, and can be shown to be loosely bounded by the norm of the pseudoinverse $\|M^\dagger\|_1$, although we will not do it here for brevity. The exact geometry strongly influences the sensitivity of the elementwise bounds, as well. Our analysis and experiments (cfr. Sections \ref{sec:gampl} and \ref{sec:i3e}) show that appreciable noise on voltages has a catastrophic impact on this style of identification. High-$\Delta^*$ identification for generic networks with a non-trivial number of branches is a hopeless task, as in this case in general a majority of branches can span a significant impedance range, and in order to form a range of solutions covering the ground truth they would have to be explored combinatorially. On the other hand, some noise in injections is tolerated. In our settings, noise on $\ell_e$ is also important, but only enters the picture downstream, as a confounder in the library refinement. The effect of noise will be investigated experimentally below.

\section{Directional sampling} \label{sec:gampl}

If direct, well-conditioned regression is not possible, we propose a technique for underdetermined identification, which outputs a range of possible solutions. The pipeline is structurally able to handle any kind of indeterminacy; however, the necessary exploration suffers from the curse of dimensionality. For sizeable feeders, it is practically applicable in cases where the exploration can be limited to a few major unidentified directions, such as those with a few degree-2 nodes.

\subsection{Bounded polytope}
We deploy robust impedance bounds derived from the library to convert polyhedron $\mathcal{Z}$ into polytope $\mathcal{Z}^\#$, limiting any unbounded direction to physical solutions and allowing its practical sampling. We add disequations to (\ref{eq:halfspacerep}) encoding global elementwise limits $r_{\mathrm{lo}} \leq \mathbf{r}\leq r_{\mathrm{hi}}$ and $x_{\mathrm{lo}} \leq \mathbf{x}\leq x_{\mathrm{hi}}$, plus line delimitations in the r-x plane $\mathbf{x} \leq m_{\mathrm{hi}}\mathbf{r}+b_{\mathrm{hi}}$, $\mathbf{x} \geq m_{\mathrm{lo}}\mathbf{r}+b_{\mathrm{lo}}$  (see also Figure \ref{fig:i3e_catalog}):

\begin{equation}
    M^\#\mathbf{z} = 
    \begin{bmatrix} 
    M \\
    \begin{bmatrix} 
    I  &0 \\
    0  &I \\
        -I & 0 \\
        0 & -I \\
        -m_{\mathrm{hi}} & I \\
        m_{\mathrm{lo}}& -I
    \end{bmatrix} 
    \end{bmatrix} \mathbf{z} 
    \leq
    d^\# = \begin{bmatrix} 
    d \\
    \ell r_{\mathrm{hi}} \\
    \ell x_{\mathrm{hi}} \\
    \ell r_{\mathrm{lo}} \\
    \ell x_{\mathrm{lo}} \\
     \ell b_{\mathrm{hi}} \\
     \ell b_{\mathrm{lo}}
    \end{bmatrix}
    \label{eq:msharp}
\end{equation}

To obtain the final samplable polytope, we select a subset of free directions, corresponding conceptually to the non-identifiable branches. This might be done by spectral analysis of $M^\#$ (whose zero singular values reveal the ambiguous directions) but the process needs careful consideration in the light of Section \ref{sec:geometry}. In particular, we underline the case where $M^\#$ experiences rank deficiency due to data solvable by a r-x prior. This happens in the test case that we will see in Section \ref{sec:i3e}, where we have a combination of 11 ambiguous edges and a costant power factor of $0.95$, both of which collapse the rank of $M^\#$; yet, the latter problem is solved by the implicit r-x prior, and the free directions are limited to the ambiguous edges.

If we split $M^\#$ according to the selected free and fixed directions, we obtain a simplified polytope by bringing the fixed portion to the RHS:

\begin{equation}
    M^\#_{\mathrm{free}} \mathbf{z}_{\mathrm{free}} \leq d^\# - M^\#_{\mathrm{fixed}} \mathbf{z}^0_{\mathrm{fixed}} = d^\#_{\mathrm{free}}
\end{equation}

Notice that $M^\#_{\mathrm{free}}$ typically has many rows of zeros; the system can be trivially simplified by removing them and the corresponding entries of $d^\#_{\mathrm{free}}$.

\subsection{Sampling}

The problem of obtaining uniform samples from high-dimensional convex polytopes arises in many different computational tasks, is challenging, and has been extensively tackled in the computational geometry literature \cite{Dyer1991-jo, Narayanan2016-wo, Chen2018-wg}. We adopt a state-of-the-art polytope sampling pipeline, consisting in preprocessing with the \textsc{Polyround} library \cite{Theorell2022-lt}, which performs simplification (removal of non-active disequations), transformation (embedding in a space where the polytope has non-zero volume) and rounding (refactoring the polytope so its covariance matrix is close to the identity matrix), followed by sampling with the Dikin walk implementation of the \textsc{PolytopeWalk} library \cite{Sun2024-lb}. To further enhance the distribution, the samples are thinned by a factor of $3$. The samples of $\mathbf{z}_{\mathrm{free}}$, naturally, include only the explored unidentifiable directions. To obtain a full-dimensional sample, we need to fill the values for the identified ones, which are uniformly set to the corresponding $\mathbf{z}^0_{\mathrm{fixed}}$ entries. A number $m$ of samples are drawn, collected in matrix $B \in \mathbb{R}^{m \times 2|\mathcal{E}|}$.

\section{Refinement}\label{sec:refinement}

In this stage, we attempt to refine the candidates collected in $B$ by ``pushing'' their entries further towards the library, according to an opportune distance function. The objective is built around the following function, which is the sum of the distances of the entries of $\mathbf{z}$ to their respective nearest library value. If we call $z_e =  \mathbf{r}_e+i\mathbf{x}_e$:

\begin{align}
        Q(\mathbf{z}) &= \sum_e\min_{c \in \mathbf{c}_e} |z_e - \ell _e c |  \\
    \nabla Q(\mathbf{z}) 
    &= \bigg[\mathrm{sign}\big( z_e - \ell_e \arg\min_{c \in \mathbf{c}_e} |z_e - \ell _e c| \big) \bigg]_{e \in \mathcal{E}}
\end{align}

We adopt a gradient descent approach with fixed learning rate $\lambda$ augmented by a penalty term to promote permanence of the solutions in $\mathcal{Z^\#}$, modulated by a coefficient $\rho$:

\begin{align}
        O(\mathbf{z}) &= Q(\mathbf{z}) + \frac{\rho}{2}\,\|[M^\#\mathbf{z}-d^\#]_{+}\|_2^2\\
    \nabla O(\mathbf{z})
    &= \nabla Q(\mathbf{z}) + \rho\,M^{\#\top}[M^\#\mathbf{z}-d
    ^\#]_{+}
\end{align}

\noindent where $[\cdot]_+ = \max(0,\cdot)$. If the library priors and length estimates are trustworthy, allowing the solution to slightly drift off $\mathcal{Z}^{\#}$ with a low $\rho$ might erase some inaccuracy inherent to the LinDistFlow approximation, yielding results closer to the ground truth. On the other hand, notice that the penalty term alone differentiates the process from a simple, elementwise snap to the nearest catalog impedance. After performing the GD, we have a collection of refined candidate solutions $C_{h, \cdot}$ in the rows of $C$.

\section{Optional thinning}\label{sec:thinning}


Having revealed the solution space with a sufficiently big number of samples, in the end one might wish to consider just a few solutions in a real study. A natural merit parameter for ranking them would be the sum of distances of each branch to the nearest library entry. However, due to the attraction exerted by the library prior during refinement, solutions will have a tendency to coalesce, and the final collection might be redundant, with many analogous high-ranking solutions. For this reason, we propose a better method to thin out $C$, if desired: that is, selecting a maximally representative collection of $m'\ll m$ samples, collected in the rows of matrix $Z$. If, for convenience, we label $H=\{1,\dots,m\}$, we can write:

\begin{equation}
    Z = C_{h' \in H', \cdot} \quad \text{with} \ \ \begin{cases}  H' \subset H \\ |H'| = m'
    \end{cases}
\end{equation}

The idea is a direct analog to what is known in the Operations Research literature as Facility Location, and in particular to the optimization of a \textit{Facility Location set function} \cite{Mirchandani1990-pc}:
\begin{equation}
\begin{aligned}
    H' = \argmin_{\tilde{H} \subset H} \quad & \bigg(  \sum_{h \in H} \min_{h' \in \tilde{H}} s(C_{h,\cdot},\ C_{h',\cdot}) \bigg) \\
    \textrm{s.t.} \quad
     & |\tilde{H}| = m'
\end{aligned}
\end{equation}

\noindent where $s(\cdot, \cdot)$ is a dissimilarity metric, for which we adopt the euclidean distance. The FL is submodular, and this property allows to tackle effectively its approximation even if this family of combinatorial problems is in general intractable \cite{Bach2013-gv}. Further acceleration is granted by approximating the pairwise similarity metric with a weighted $K$-neighbors graph \cite{Wei2014-xh}.

\section{GPU parallelization} \label{sec:gpuimpl}

GPU acceleration can be leveraged in the computations and was important in achieving good wall-clock performance.

\begin{enumerate}[leftmargin=*]

    \item The gradient descent is well-suited for GPU acceleration. The computation of $\nabla Q$ is overtly elementwise and was packaged in a kernel written in \textsc{Numba-CUDA}\footnote{https://nvidia.github.io/numba-cuda/}. The resulting $C'$ representation in GPU memory can be cast directly to \textsc{Cupy} arrays \cite{Okuta2017-my}, allowing acceleration of the subsequent matricial computations with no transfer overhead. The entire GD step, and therefore the entire loop, is realized on-device.

    \item The facility location submodular optimization of Section \ref{sec:thinning} is run with the \textsc{Apricot} library \cite{Schreiber2019-kw} on a precomputed similarity graph instead of the raw data, which requires the computation of a weighted $K$-neighbors graph over $C$ (which is already in device memory at the end of the PGD). This task is accelerated on GPU with the \textsc{cuML} library \cite{Raschka2020-yc}.
\end{enumerate}

\noindent The average speedup across 10 runs are shown in Table \ref{tab:gpuspeedup}, divided by accelerated task. In the table, the GD step excludes the computation of $\nabla Q$, which is evaluated separately. The GD step is evaluated with and without the penalty term gradient computation, which is the heaviest part.

\begin{table}[h!]				
\caption{GPU speedup (average across 10 runs).}				
\centering				
\begin{tabular}{|l|c|c|}				
\hline				
\rowcolor[HTML]{EFEFEF} 				
\multicolumn{1}{|c|}{\cellcolor[HTML]{EFEFEF}\textbf{Task}}  &	\textbf{CPU}	&	\textbf{GPU}	\\  \hline
$\nabla Q$            &	0.13920	s &	0.00058	s \\  \hline
GD step ($\rho=0$)        &	0.01817	s &	0.00479	s \\  \hline
GD step ($\rho>0$)        &	0.16886	s &	0.03029	s \\  \hline
KNN fit-transform    &	4.50080	s &	0.21112	s \\  \hline
\end{tabular}				
\label{tab:gpuspeedup}				
\end{table}

\section{Application to standard test case}\label{sec:i3e}

\begin{figure}[hbt!]
    \centering
    \includegraphics[trim={0cm 0.7cm 0cm 0.1cm},width=.95\columnwidth]{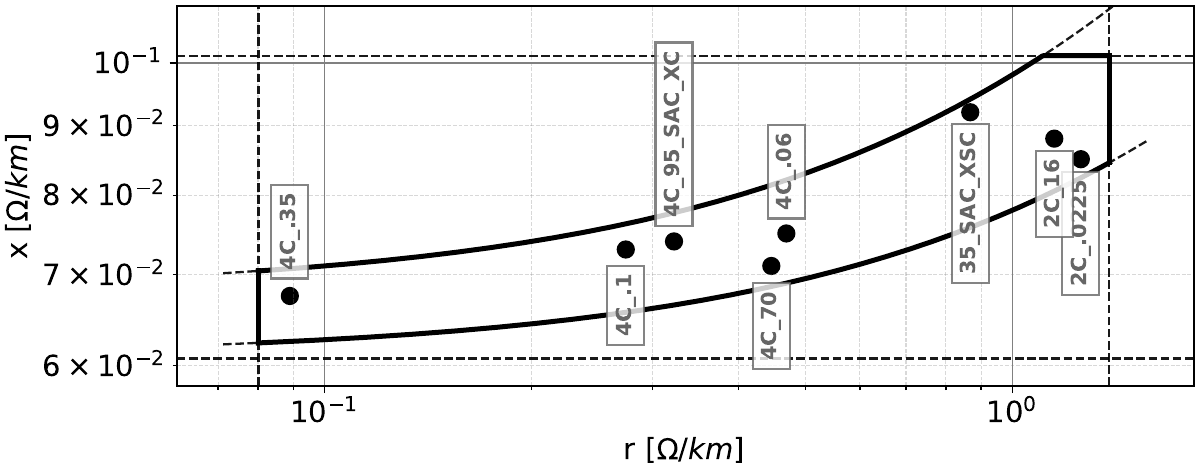}
    \caption{Library of the IEEE LV test network, with bounds from (\ref{eq:msharp}). See Table \ref{tab:workingparams}. The graph is in log scale on both axes.}
    \label{fig:i3e_catalog}
\end{figure}

We test our method on the IEEE European Low Voltage Feeder dataset. We used the reduced model presented in \cite{Khan2022-ls}, which is electrically equivalent to the original, but discards many buses irrelevant from a topological and electrical viewpoint, included in the original representation for plotting purposes. This was also done in \cite{Guo2025-hp}. This network has 116 nodes, of which 6 are of degree 2. The dataset includes 1440 injection datapoints. The injection snapshots all have a fixed power factor of $0.95$ at all loads, making the library prior important for tight identification according to our analysis in Section \ref{sec:geometry}, as we will see shortly. The loads were equally distributed to all three phases to obtain a balanced network (we aim to expand the tool to unbalanced operating conditions in future work). The network is simulated in full AC power flow with the Pandapower package \cite{Thurner2018-ts}, and only the relevant simulated results at the leaves are retained, replicating smart meter data. The library prior at all branches pessimistically includes all the wire types present in the network (see Figure \ref{fig:i3e_catalog}).
The parameters used for the analysis of this section are summarized in table \ref{tab:workingparams}. The gradient descent learning rate was chosen conservatively with respect to the experienced convergence. The whole $m$ range is shown in the identification results, skipping the optional thinning step. The best results were obtained by not using the penalty term governed by $\rho$, as the $\mathcal{Z}^\#$ samples were already of high quality, and the only way to further enhance them was to let the prior, in our case exact, erase the inherent linearization inaccuracies in many of the branches.

\begin{table}[h!]
\caption{Parameter settings.}
\centering
\begin{tabular}[t]{|c|l|}
\hline
\rowcolor[HTML]{EFEFEF} 
{\color[HTML]{000000} \textbf{Symbol}} & {\color[HTML]{000000} \textbf{Value}} \\ \hline

$\kappa$                               & $1.05$                            \\ \hline
$\lambda$                                    & $0.01$         \\ \hline
$\rho$                                 & $0$                                 \\ \hline
$m$                                    & 30 000                                  \\ \hline
$m'$                                   & ---                                  \\ \hline
$K$                                    & ---                                   \\ \hline
\end{tabular}
\begin{tabular}[t]{|c|l|}
\hline
\rowcolor[HTML]{EFEFEF} 
{\color[HTML]{000000} \textbf{Symbol}} & {\color[HTML]{000000} \textbf{Value}} \\ \hline
$r_\mathrm{hi}$   & $1.10\times r_{\mathrm{max}}$  \\ \hline
$x_\mathrm{hi}$   & $1.10\times x_{\mathrm{max}}$  \\ \hline
$r_\mathrm{lo}$   & $0.90\times r_{\mathrm{min}}$           \\ \hline
$x_\mathrm{lo}$   & $0.90\times x_{\mathrm{min}}$                            \\ \hline
$m_\mathrm{hi}$   & 0.030  \\ \hline
$b_\mathrm{hi}$   & 0.068                                  \\ \hline
$m_\mathrm{lo}$   & 0.017                                   \\ \hline
$b_\mathrm{lo}$   & 0.061                            \\ \hline
\end{tabular}
\label{tab:workingparams}
\end{table}

\subsection{Identification results}

\begin{figure}[hbt!]
    \centering    
    \includegraphics[width=\columnwidth]{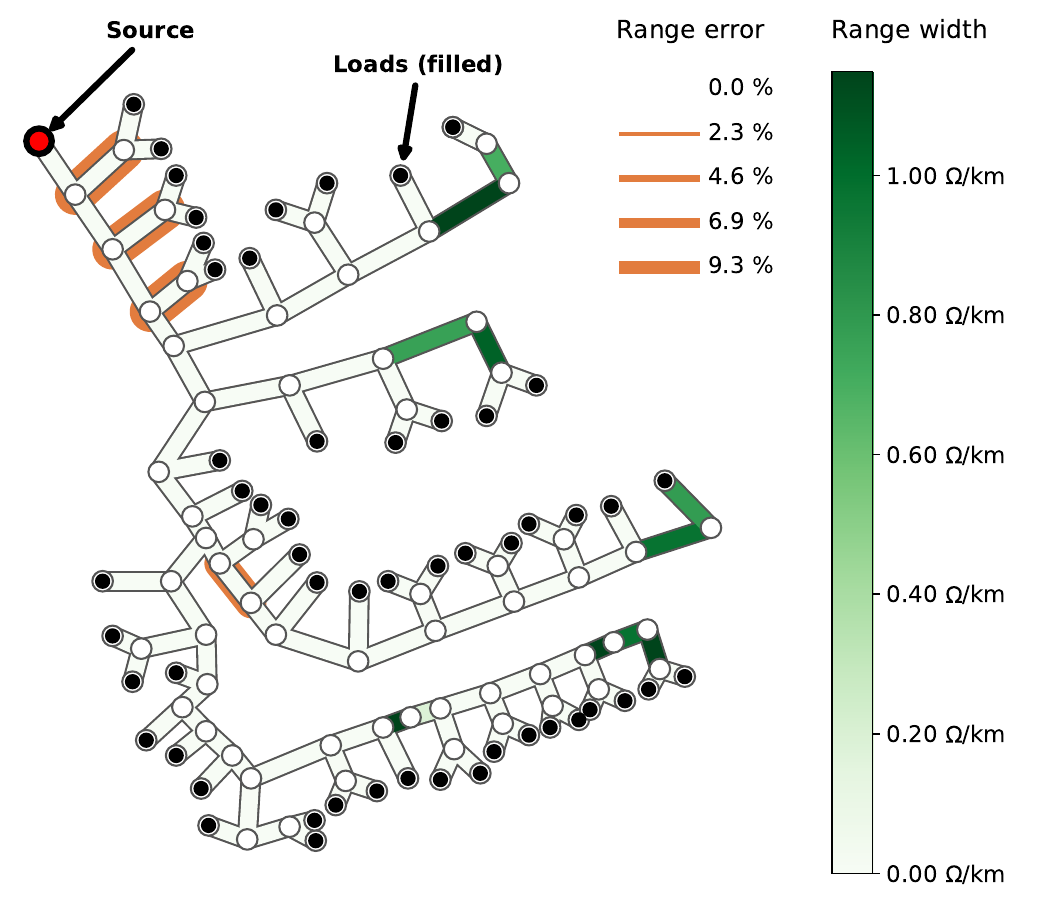}
    \caption{Identification results}
    \label{fig:id1}
\end{figure}

An example of individual identification is performed with a set of just $10$ randomly selected datapoints. In principle, noiseless identification becomes possible as soon as system (\ref{eq:halfspacerep}) becomes determined \cite{Grammatikos2025-mx}; we found that the performance of the tool essentially saturates with a few datapoints. The selected free directions correspond to the $11$ edges that include a degree-2 node, which are consequently the only one with a range $>0$. The results are summarized in Figure \ref{fig:id1}, which visualizes the range of offered solutions, in terms of impedance magnitude, marginalized at each edge. If the ground truth is out of the range, this is indicated as an orange border of thickness proportional to the entity of the error; notice that this happens only for a few non-ambiguous branches, and is due, to our best understanding, to the LinDistFlow approximation.

Due to the underdetermined and range-based nature of the presented method, a direct comparison with existing methods is not straightforward. We argue that the core objective is the capture of the ground truth solution in a range which is as tight as possible. If we use the MAPE of the closest solution in the range, that is:

\begin{equation}
    \begin{cases}
\mathrm{MAPE}^{R}_*
= \min_h \bigg(
\frac{1}{n}
\sum_{i=1}^{n}
\left|
\frac{\Re(Z_{h,i})-\Re(z^{\mathrm{true}}_i)}{\Re(z^{\mathrm{true}}_i)}
\right|\bigg)
\\[1.2em]
\mathrm{MAPE}^{X}_*
= \min_h \bigg(
\frac{1}{n}
\sum_{i=1}^{n}
\left|
\frac{\Im(Z_{h,i})-\Im(z^{\mathrm{true}}_i)}{\Im(z^{\mathrm{true}}_i)} 
\right| \bigg)
\end{cases}
\end{equation}

The above identification attains $\mathrm{MAPE}^{R}_*
= 2.25\%$ ($4.73\%$ before refinement) and $\mathrm{MAPE}^{X}_*
= 0.33\%$ ($7.72\%$ before refinement), which are superior to the corresponding IEEE European Low Voltage Feeder identification in \cite{Guo2025-hp}, even if it employs a larger dataset of size 50 and the removal of problematic degree-2 nodes.

\begin{figure}[hbt!]
    \centering    
    \includegraphics[width=\columnwidth]{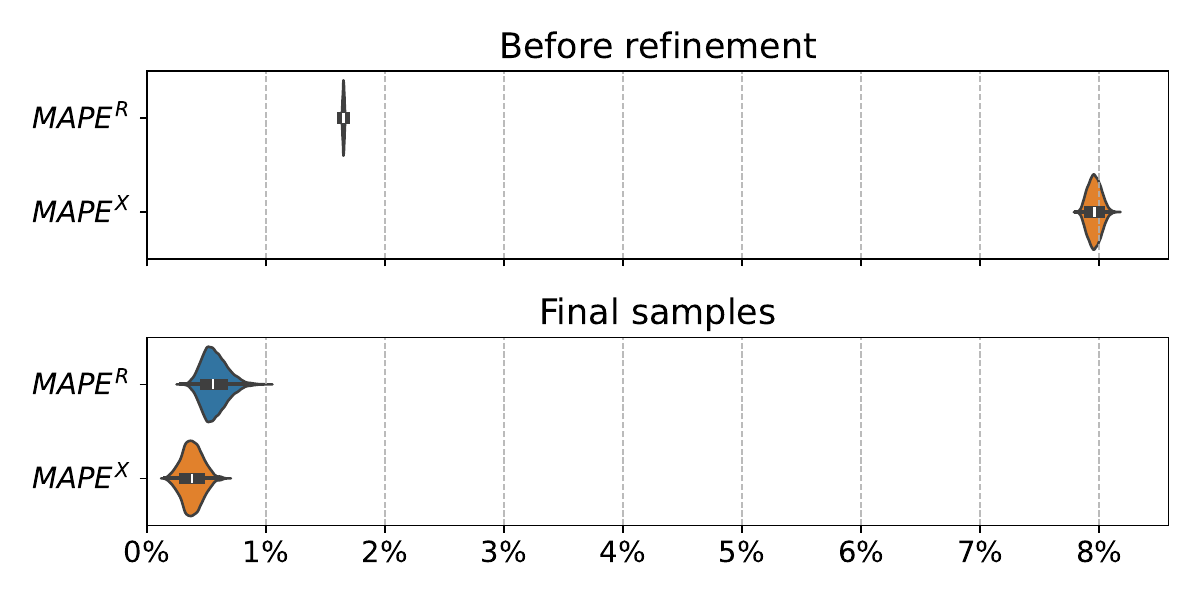}
    \caption{Violin plot of the MAPE values mapped to the degree-simplified network setup.}
    \label{fig:simpmape}
\end{figure}

For the sake of closer comparability with the mentioned literature background, we can map the results onto the corresponding full-rank identification by summing the impedances of the chains of ambiguous edges. This reduces to the pure regression setting for the simplified network. The results of this study, performed on 50 datapoints, are shown in Figure \ref{fig:simpmape}, in which we also show the same quantity before performing the refinement. The method shows a high performance on the equivalent simplified task, and it can be clearly appreciated how the prior has a strong effect in adjusting the r/x ratio of the samples, which, as anticipated, is beneficial in this test case due to the fixed high power factors impairing the identification.

We wish to stress that error is not evenly distributed across branches: the median percentage error of the individual solutions in the range is negligible, with the mean absolute error being driven by extreme values due to the sporadic misattributions.

\section{Tolerance to noisy input} \label{sec:noise}

\begin{figure}[hbt!]
    \centering    
    \includegraphics[width=.9\columnwidth]{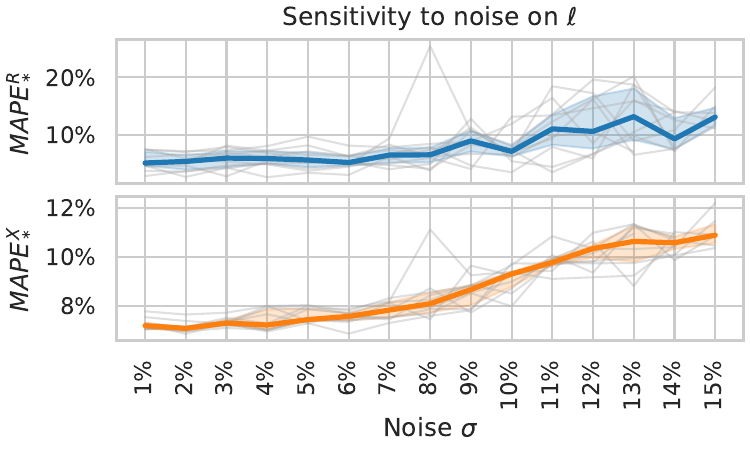}
    \caption{Sensitivity to noise on lengths. The line estimator is the median with $50^\mathrm{th}$ percentile interval.}
    \label{fig:noise_10}
\end{figure}

\begin{figure}[hbt!]
    \centering    
    \includegraphics[width=.9\columnwidth]{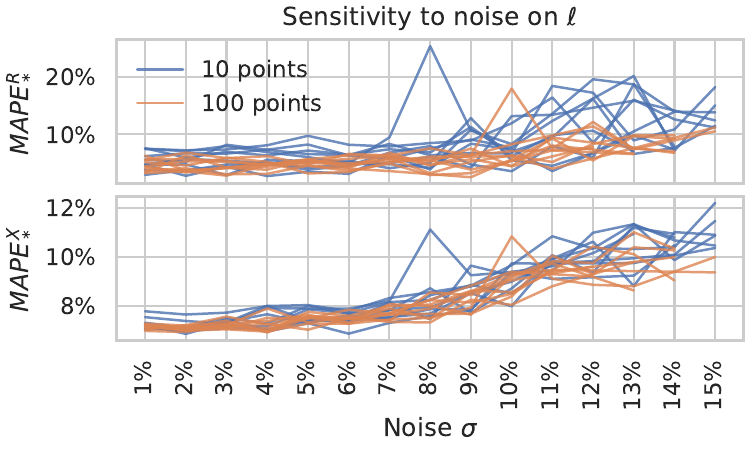}
    \caption{Effect of larger dataset on the $MAPE_*$ in the presence of $\ell$ noise.}
    \label{fig:noise_100}
\end{figure}

Using the library as a prior, as opposed to a set of hard combinatorial choices, allows the method to handle naturally imprecision in the length data $\ell_e$, which in real studies would only be known approximately. We tested gaussian noise with different variances, across $10$ different datapoint collections. An analysis of the effect is presented in Figure \ref{fig:noise_10}. In this case, $\rho$ was set to $0.05$, as the noise on lengths obfuscates the precision of the prior, and data compatibility is needed to avoid major non-physical drifts. In the presence of noise on $\ell$, a higher dataset helps stabilizing the results. In Figure \ref{fig:noise_100}, the effect of using $100$ datapoints is shown, attaining MAPE at the lowest end of the $10$-datapoint spectrum with less spread.

\begin{figure}[hbt!]
    \centering    
    \includegraphics[width=.9\columnwidth]{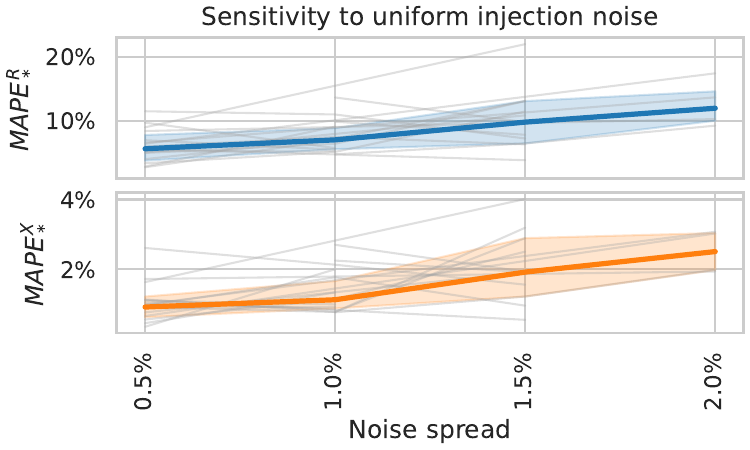}
    \caption{Effect of injection noise on the $MAPE_*$. The line estimator is the median with $50^\mathrm{th}$ percentile interval.}
    \label{fig:noise_inj}
\end{figure}

We also analyzed the sensitivity to uniform injection noise, obtaining harsher results. The data incoherency often results in infeasible sets of constraints, particularly as the noise spread and the number of datapoints increase; in these cases, plain regression would be the standard alternative. The results computed on the valid runs of $10$ datapoints are shown in Figure \ref{fig:noise_inj}, but these values should be evaluated taking survivorship bias into consideration.

\begin{figure}[hbt!]
    \centering    
    \includegraphics[width=\columnwidth]{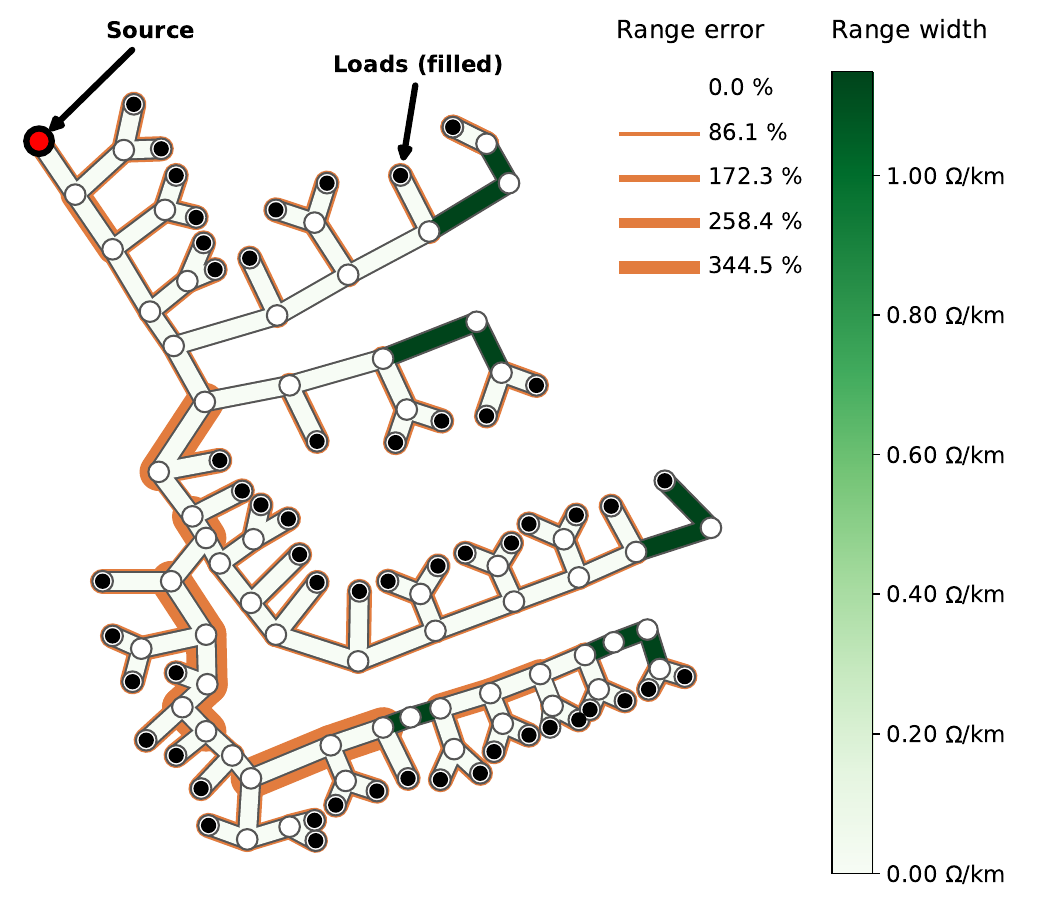}
    \caption{Identification fails for normal voltage noise with $\sigma=0.5\%$ (200 datapoints).}
    \label{fig:idvnoise}
\end{figure}

Noise on voltage, as observed above, directly inflates $\Delta^*$ and immediately overcomes the fine relationships among the terminal voltage magnitudes that, through simultaneous voltage drop equations, restrict to a usable sliver those at the latent nodes, which is ultimately what makes each branch identifiable. Normal noise with a variance of $0.5\%$ is sufficient to derail the identification, even with many datapoints; see Figure \ref{fig:idvnoise}.

\section{Integration in GIS-based identification workflow}\label{sec:gis}

Besides the ability to directly address test cases such as the one above without simplifying them, we argue that the approach lends itself to topology identification workflows aiming to synergize smart meter measurements and GIS asset databases. The value of the information kept in GIS systems by the DSOs is recognized in the literature; however, its completeness is sufficient for asset management purposes but not for trustworthy and immediate connectivity reconstruction \cite{Navarro2015-jm, Luan2015-tv, Shahid2019-mk}. Therefore, the available topological information can be distilled into spatially-referenced nodes (cabinets, joints, delivery points) with underlying connectivity that must be inferred. In this setup, candidate connections with degree-2 might arise in general and reflect the reality on the field. The approach presented in this paper allows to constrain proposals to smart meter data and validate them against a known conductor library, thus exploiting the combined value of these heterogeneous sources of information, which are often the only ones available. A fully integrated approach of this nature will be detailed in a forthcoming publication.

\section{Conclusion and future work}\label{sec:conclusion}

In this paper, we offered an analysis of impedance identifiability based on a feasible impedance polyhedron, which offers a unifying view of various issues and induces an impedance estimation method. Compared to the current literature, the method is able to handle intrinsically ambiguous networks and outputs a range of solutions by revealing the algebraic structure of impedance solutions compatible with the available data. We use a conductor library to bound, orient and refine solutions, a relevant piece of data which can be often be obtained in real studies and has seen limited use. A few operating parameters have to be set, but they have intuitive interpretations and none was found to have an ambiguous or volatile impact on the final solution. Very good identification accuracy in a sizeable case study was attained, as well as reasonable resistance to noise on the length and injection data. The solutions carried slight relative imprecision due to the underlying LinDistFlow approximation, but its magnitude is manageable and limited to a minor number of branches, with global MAPE below regression methods that address the same setting, with less data and without removal of degree-2 nodes.
There are several avenues of research that can be further explored starting from the basis of this work. Naturally, the first one is connectivity reconstruction with combined GIS data, which was outlined in Section \ref{sec:gis}. Furthermore, the authors identify and intend to pursue the following:

\begin{enumerate}[leftmargin=*]
    \item The method might be applicable to three-phase unbalanced networks by performing opportune network transforms and considering one phase at a time, at least in the perfectly transposed case; this idea needs closer investigation.
    \item LinDistFlow is the linearization of the power flow manifold obtained around one specific point: the no-load condition \cite{Bolognani2015-sp}. The method could be rebuilt around a different linearized model, such as \cite{Yang2017-it}, and the performance impact assessed. An interesting generalization would be an attempt to re-assemble $\mathcal{Z}$ iteratively under equations linearized online around the current best estimate of the true operating point. The key result would be finding under which conditions, if at all, such a process converges to the underlying true solution.
\end{enumerate}

\section*{Acknowledgment}
This paper has been supported by the FEDECOM project, under the European Union’s Horizon Europe programme with Grant Agreement No. 101075660.
The authors would like to thank Dr. Roberto Rocchetta for the valuable discussions and insightful comments throughout the development of this work.

\bibliographystyle{IEEEtran}
\bibliography{paperpile}

\begin{IEEEbiography}[{\includegraphics[width=1in,height=1.25in,clip,keepaspectratio]{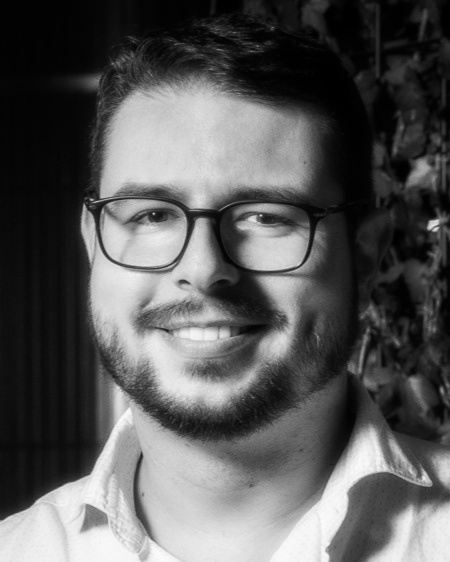}}]{Federico Rosato}
received the B.Sc. and M.Sc. degrees in Energy Engineering (2011, 2014) from Politecnico di Milano, Milano, Italy, and the Ph.D. degree in Electrical Engineering (2023) from Politecnico di Milano, Milano, Italy.

He is currently a Researcher with the University of Applied Sciences and Arts of Southern Switzerland (SUPSI). His research focuses on data analytics, control and planning for power systems.
\end{IEEEbiography}

\begin{IEEEbiography}[{\includegraphics[width=1in,height=1.25in,clip,keepaspectratio]{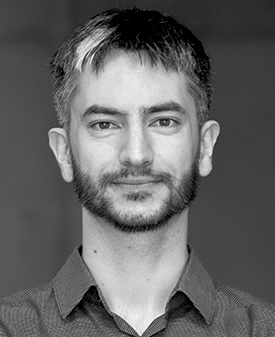}}]{Lorenzo Nespoli}
Received the M.Sc. in Energy Engineering in 2014 from Politecnico di Milano, Milano, Italy and the Ph.D. from École Doctorale in Energy (EDEY) from Swiss Federal Institute of Technology Lausanne (EPFL), Lausanne, Switzerland, in 2019. He is leading the Smart Energy Systems team at SUPSI, where he’s a Senior Researcher, focusing on forecasting, control and decision making under uncertainty in energy systems and infrastructure. 
\end{IEEEbiography}

\begin{IEEEbiography}[{\includegraphics[width=1in,height=1.25in,clip,keepaspectratio]{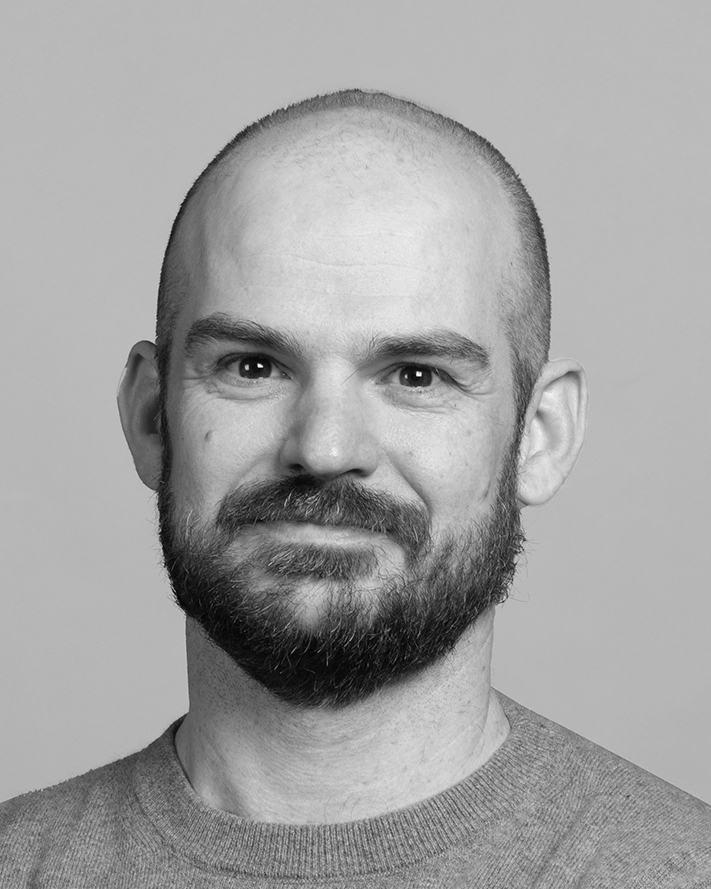}}]{Vasco Medici}
received the M.Sc. degree in Microengineering from the Swiss Federal Institute of Technology Lausanne (EPFL), Lausanne, Switzerland, in 2005, and the Ph.D. degree in Neuroinformatics from ETH Zurich, Zurich, Switzerland, in 2010. He is currently an Associate Professor in Intelligent Energy Systems and Head of the Energy Systems Sector with the Institute for Applied Sustainability to the Built Environment (ISAAC), University of Applied Sciences and Arts of Southern Switzerland (SUPSI). His research focuses on intelligent energy systems, smart grids, and data-driven methods for power and energy systems.
\end{IEEEbiography}

\vfill

\end{document}